\pdfoutput=1
\documentclass{article}
\usepackage{spconf,amsmath,amssymb,graphicx,booktabs,array,float,balance,microtype}
\usepackage[T1]{fontenc}
\usepackage{newtxtext,newtxmath}
\usepackage[table]{xcolor}
\usepackage{xfp}            % computes the routed gains in Table 1 automatically
\usepackage{cite}           % [7]--[13] style citation ranges
\usepackage[hidelinks]{hyperref}
\hypersetup{pdftitle={Beyond Two Tracks: Speaker-Anchored Full-Duplex Dialogue Recovery},
 pdfkeywords={full-duplex dialogue, speaker-anchored routing, dialogue recovery, speech separation}}
\title{BEYOND TWO TRACKS: SPEAKER-ANCHORED\\ FULL-DUPLEX DIALOGUE RECOVERY}
\name{
Yuxuan Chen$^{1}$ \qquad
Yang Xiao$^{2}$ \qquad
Jack Sun$^{3}$ \qquad
Chak Sun$^{3}$ \qquad
Brick Li$^{3}$ \qquad
Sam$^{3,\dagger}$
\thanks{
$^{\dagger}$Corresponding author.
}
}

\address{
$^{1}$ The Chinese University of Hong Kong, Shenzhen\\
$^{2}$ The University of Melbourne \quad
$^{3}$ Lulula\\
yuxuanchen5@link.cuhk.edu.cn,\;
yang.xiao.1@student.unimelb.edu.au\\
jack@lulula.ai,\;
chak@lulula.ai,\;
brick@lulula.ai,\;
sam@lulula.ai
}
\definecolor{ours}{RGB}{24,95,175}       % proposed method (label, best bars)
\definecolor{oursbg}{RGB}{228,237,249}   % row tint for proposed method
\definecolor{barbase}{RGB}{196,196,196}  % baseline WT bars
\definecolor{gain}{RGB}{18,122,71}       % routed improvement
\definecolor{flat}{RGB}{118,118,118}     % de-emphasised text / arrows

\newcommand{\wtbar}[2]{{\color{#2}\rule[-0.15ex]{\dimexpr #1pt*3/5\relax}{1.25ex}}}
\newcommand{\wt}[1]{\makebox[58pt][l]{\makebox[24pt][r]{#1}\hspace{3pt}\wtbar{#1}{barbase}}}
\newcommand{\wtb}[1]{\makebox[58pt][l]{\makebox[24pt][r]{\textbf{#1}}\hspace{3pt}\wtbar{#1}{ours}}} % best in group, ours
\newcommand{\pairf}[2]{\leavevmode{\color{flat}#1\,$\rightarrow$}\,#2}
\newcommand{\pairg}[2]{\leavevmode{\color{flat}#1}\,{\color{gain}$\rightarrow$}\,\textbf{\color{gain}#2}%
  \makebox[0pt][l]{\,{\fontsize{6.5}{7}\selectfont\color{gain}$+$.\fpeval{round((#2-#1)*1000,0)}}}}

\newcommand{\panel}[1]{{\color{ours}\rule[-0.3ex]{2.2pt}{1.6ex}}\hspace{4pt}\textbf{#1}}
\newcommand{\meta}[1]{\leavevmode{\color{flat}#1}}
\newlength{\lblw}\newlength{\colw}

\newcommand{\grp}[1]{\multicolumn{3}{@{}l}{\textit{#1}}}

\newcommand{\rate}[1]{\makebox[22pt][r]{#1}\hspace{3pt}%
  \makebox[20pt][l]{{\color{barbase}\rule[-0.15ex]{20pt}{1.25ex}}%
  \hspace{-20pt}{\color{ours}\rule[-0.15ex]{\dimexpr #1pt/5\relax}{1.25ex}}}}
\newlength{\lblwa}\newlength{\colwa}\newlength{\lblwc}\newlength{\colwc}
\begin{document}
\ninept
\maketitle
\begin{abstract}
Spoken interaction is moving from alternating turns to full-duplex conversation, where an agent listens while speaking. Learning this coordination requires natural dialogue with persistent speaker-resolved tracks, a data requirement that remains difficult to satisfy at scale. Generative recovery offers a route from ordinary recordings, but recovering two voices is not enough: their track assignments can drift as the conversation unfolds. We address this missing link with speaker-anchored routing before diffusion: a 1.84\,M-parameter module matches the separator's coarse streams to fixed enrollment identities and orders them before a frozen generator reconstructs the speech, without retraining the generator or altering the reconstructed audio. On 25-minute remote conversations, routing cuts wrong-track time from 43.9\% to 4.1\%, below the 6.2\% of matched-enrollment waveform re-identification, with its largest advantage when one participant holds the floor (17.8\% versus 28.6\%), and raises fixed-track activity F1 from 0.46 to 0.84 while per-chunk reconstruction quality is unchanged. The same router transfers to unseen face-to-face and telephone recordings, and a margin-gated verification step brings their wrong-track time to 0.9--1.7\% while embedding 20--36\% of the output, where full re-identification embeds all of it. Recovered voices thus become the participant-consistent tracks that full-duplex dialogue modeling needs. Audio examples:\newline{\footnotesize\url{https://elpsykongloo.github.io/beyond-two-tracks/}}
\end{abstract}
\vspace{-5pt}

\section{Introduction}
\label{sec:intro}
\looseness=-1 Speech interaction is moving beyond ``listen, then answer.'' A conversational agent should keep listening while it speaks, respond to an interruption, and distinguish a brief acknowledgment from a request to take the floor. Full-duplex spoken dialogue brings these behaviors into a continuous interaction rather than a sequence of isolated turns~\cite{chen2025survey,lu2026survey}. This changes what a model must learn: not just the content of a response, but how two participants coordinate speech and silence on a shared timeline. Full-duplex models now span parallel-stream generation~\cite{nguyen2023dgslm,defossez2024moshi}, listening-while-speaking agents~\cite{ma2024lslm,veluri2024syncllm}, language-model-based duplex systems~\cite{wang2024neuralfsm,wang2024freezeomni,chen2025minmo,zhang2025omniflatten,wang2025ntpp,hu2025salm,yu2025salmonn}, and multi-turn evaluation~\cite{wang2026covo,lin2025fdbv2}.

The modeling frontier is advancing faster than the supply of suitable conversational data. Both surveys identify the scarcity of synchronized, speaker-resolved natural dialogue as a recurring bottleneck~\cite{chen2025survey,lu2026survey}. CANDOR and Seamless Interaction expand access to natural dyadic recordings~\cite{reece2023candor,agrawal2025seamless}, while synthetic resources such as DuplexDrama target richer interaction behaviors~\cite{guo2026duplexdrama}. A complementary opportunity is to recover full-duplex tracks from the much broader pool of existing mixed recordings. Unlike transcribed turns, these recordings retain the timing of speaking, listening, and overlap; to exploit it, the recovered channels must also retain who produced it.

DialogueSidon~\cite{nakata2026dialoguesidon} enables this route by jointly restoring and separating in-the-wild dialogue; DuplexChat uses it to construct speaker-separated conversational data~\cite{nakata2026duplexchat}. Yet two recovered tracks are not yet two persistent participants. A locally successful separator may exchange speaker assignments across processing windows: the audio stays plausible while a speaker's continuation moves onto the partner's channel. On 25-minute conversations, the released overlap-stitching rule leaves 44.9\% of speech on the wrong track, no better than leaving chunks unaligned. Permutation-invariant training avoids an arbitrary source order~\cite{yu2016pit} and overlap stitching gives local continuity~\cite{yoshioka2019css}; neither fixes who each track is across a conversation.

The errors concentrate where conversations spend much of their time: one person holds the floor while the other listens. Such activity-asymmetric chunks carry 52.2\% of speech-bearing duration on CANDOR, and there both overlap evidence and reconstructed-waveform evidence weaken, because one track is nearly silent. Current recovery pipelines assign identity after reconstruction: ConversationalVoice maps recovered slots using diarization, activity, and speaker evidence~\cite{he2026conversationalvoice}, and enrolled waveform re-identification scores already generated tracks; speaker-conditioned extraction instead changes signal estimation itself~\cite{wang2019voicefilter}. We instead make participant identity an input to the existing generation process.

We propose \textbf{speaker-anchored routing before diffusion}: match the separator's coarse streams to fixed enrollment anchors, choose their order, and let the frozen generator reconstruct the speech. This separates who belongs on each track from how the speech is reconstructed. Our contributions are: (i) persistent participant control with a frozen generator, cutting CANDOR wrong-track time from 43.9\% to 4.1\% with 1.84\,M parameters and 2.8\% added runtime; (ii) activity-stratified evidence that locates the advantage over matched-enrollment re-identification in floor-holding speech (17.8\% versus 28.6\%); and (iii) a routing margin that directs verification to 20--36\% of the output, together with interventions showing why the pre-diffusion interface works.

\section{Speaker-anchored recovery}
\label{sec:method}
\subsection{Assign participants before generating their tracks}
\begin{figure*}[t]
\centering
\includegraphics[width=\textwidth]{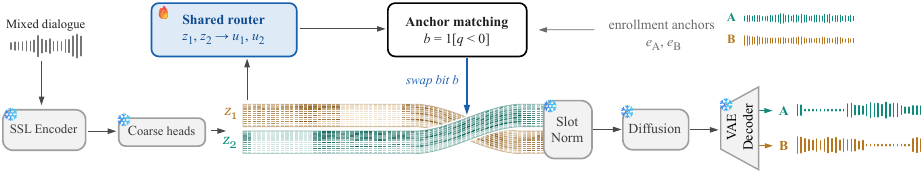}
\caption{\textbf{A persistent identity interface to a frozen generator.} Snowflakes mark the released modules, kept exactly as published; the flame marks the only trained one, the 1.84\,M-parameter shared router, and parameter-free anchor matching carries neither. The two coarse streams (32-D, 50\,Hz) are drawn as cell grids and keep their colour through to their output track. The router pools each stream into an embedding $u_i$, matching compares $u_1-u_2$ against the anchors $e_A-e_B$---the crossing pairing wins here---and the swap bit $b$ exchanges the two slots, so the grids reach the generator with their values and timing untouched. Not drawn: the mixture features that bypass the router, and the optional verification of the decoded tracks when $m<\theta$.}
\label{fig:pipeline}
\end{figure*}
The inputs are a mixed two-person recording and enrollment speech for participants A and B; the requested output order is fixed throughout. The frozen DialogueSidon encoder and coarse heads produce two 32-dimensional coarse streams at 50\,Hz and mixture features. A shared router maps each stream to a unit 192-dimensional speaker embedding $u_1,u_2$. Normalized enrollment anchors $e_A,e_B$ come from a frozen ECAPA-TDNN encoder~\cite{desplanques2020ecapa,ravanelli2021speechbrain}. Assignment reduces to
\begin{equation}
q=\langle u_1-u_2,e_A-e_B\rangle,\quad b=\mathbf{1}[q<0],\quad m=|q|.
\label{eq:route}
\end{equation}
The bit $b$ selects the direct or exchanged order, and $m$ records the margin. Complete streams are permuted before the released slot normalization: their values and temporal positions are retained. Mixture features bypass the router, and the frozen diffusion model and VAE decoder reconstruct the speech (Fig.~\ref{fig:pipeline}).

\subsection{Learn speaker matching, retain reconstruction}
Three temporal convolutions of width 384 and dilations 1, 2, and 4 feed learned speech gating, attentive statistics pooling~\cite{okabe2018pooling}, and an embedding projection. Both streams share weights and input normalization. The router contains 1.84\,M trainable parameters and adds 2.8\% to the median end-to-end real-time factor (0.0193 $\rightarrow$ 0.0198 on a 20-session timing subset, excluding optional verification); all released separator weights remain frozen.

Training combines in-batch contrastive matching, two-way anchor cross-entropy, and activity binary cross-entropy weighted by 0.3. Alternating steps use clean target latents (with reference activity and activity supervision) or coarse predictions from degraded mixtures with learned activity weights. With \emph{swap inheritance}, a chunk whose margin is below 0.05 keeps the preceding swap bit, so a silent listener cannot reset the convention; the first chunk uses its own decision.

Optional verification embeds the two already decoded tracks when $m<\theta$, with default $\theta=0.30$. It changes assignment when the exchanged enrollment score exceeds the direct score by 0.10. Verification adds selective speaker-encoder computation, not another diffusion or decoding pass.

\section{Experimental setup}
\label{sec:setup}
\textbf{Data and training.} The main router uses 500 CANDOR training sessions and 40 development sessions~\cite{reece2023candor}; splits are session-disjoint. The 29,958 degraded 20\,s examples use reverberation, MUSAN noise/music~\cite{snyder2015musan}, band limitation, clipping, MP3 coding, and packet loss. We use AdamW for 15,000 steps, batch 48, peak learning rate $10^{-3}$, and contrastive temperature 0.07; development order accuracy selects the checkpoint.

Evaluation covers 624 real dialogues (62.9\,h; Table~\ref{tab:main}): 64 held-out CANDOR sessions truncated to 25\,min; 360 Seamless Interaction improvised dyads (48\,kHz)~\cite{agrawal2025seamless}; and the first 200 SpokenWOZ test dialogues by ID (8\,kHz), without further filtering~\cite{si2023spokenwoz}. Seamless keeps dyads of at least 1.5\,min with both participants voiced for at least 8\% of the recording, minus one near-duplicate pair (anchor cosine above 0.85). Seamless and SpokenWOZ are unseen in router training and test transfer to new speakers and recording conditions. Evaluation mixtures are equal-weight sums of reference channels without added degradation.

\begin{table*}[t]
\centering
\caption{\textbf{Identity across contexts and recording domains, with acoustic quality retained.}
(a) Duration-pooled wrong-track time (WT, \%\,$\downarrow$), grouped by recognition cost: the lower group runs a speaker encoder on the decoded tracks, and the percentage after a label is the share of output it embeds. Bars scale with WT; blue marks the proposed method.
(b) Unrouted $\rightarrow$ routed acoustic scores ($\uparrow$): ``best'' allows a better permutation per chunk, ``fixed'' keeps the participant convention, and bold marks the routed gain; all baselines share the unrouted generation.}
\label{tab:main}
\begingroup
\fontsize{8.3}{9.3}\selectfont
\setlength{\aboverulesep}{0.25ex}\setlength{\belowrulesep}{0.4ex}
\setlength{\tabcolsep}{5pt}
% Fixed equal column widths (instead of tabular* + \extracolsep) so \rowcolor covers the whole row.
\settowidth{\lblw}{\textbf{Routing + verification} \meta{20--36\%}}
\setlength{\colw}{\dimexpr(\textwidth-\lblw-40pt)/4\relax} % 40pt = the 8 interior \tabcolsep
\begin{tabular}{@{}l *{4}{>{\centering\arraybackslash}p{\colw}}@{}}
\toprule
 & \multicolumn{2}{c}{\textbf{CANDOR: long remote dialogue}} & \textbf{Seamless: face-to-face} & \textbf{SpokenWOZ: telephone}\\
\cmidrule(lr){2-3}\cmidrule(lr){4-4}\cmidrule(l){5-5}
\meta{Chunk / overlap; sessions / h} & \meta{30 / 3\,s; 64 / 26.67} & \meta{120 / 10\,s; 64 / 26.67} & \meta{30 / 3\,s; 360 / 25.28} & \meta{30 / 3\,s; 200 / 10.91}\\
\midrule
\multicolumn{5}{@{}l}{\panel{(a) Participant assignment: WT (\%)}\quad\meta{\itshape grouped by recognition cost on the decoded tracks}}\\
No alignment            & \wt{43.86} & \wt{18.79} & \wt{18.66} & \wt{13.39}\\
Released overlap cosine & \wt{44.89} & \wt{21.23} & \wt{18.35} & \wt{12.82}\\
\rowcolor{oursbg}
\textbf{\color{ours}Pre-diffusion routing} & \wtb{4.08} & \wtb{1.89} & \wtb{4.17} & \wtb{2.32}\\
\addlinespace[2pt]
Enrolled waveform re-ID \meta{100\%} & \wt{6.20}  & \wt{4.72}  & \wt{2.15} & \wt{0.57}\\
\rowcolor{oursbg}
\textbf{\color{ours}Routing + verification} \meta{20--36\%} & \meta{---} & \meta{---} & \wtb{1.74} & \wtb{0.87}\\
\midrule
\multicolumn{5}{@{}l}{\panel{(b) Reconstruction versus ownership: unrouted $\rightarrow$ routed}}\\
Speaker similarity, best  & \pairf{.435}{.438} & \pairf{.507}{.509} & \pairf{.559}{.558} & \pairf{.625}{.621}\\
Speaker similarity, fixed & \pairg{.268}{.427} & \pairg{.313}{.504} & \pairg{.366}{.546} & \pairg{.458}{.616}\\
Activity F1, best         & \pairf{.849}{.851} & \pairf{.850}{.852} & \pairf{.868}{.870} & \pairf{.926}{.923}\\
Activity F1, fixed        & \pairg{.464}{.839} & \pairg{.471}{.851} & \pairg{.546}{.853} & \pairg{.424}{.911}\\
\bottomrule
\end{tabular}
\par\vspace{2pt}
\begin{minipage}{\textwidth}\fontsize{7.6}{8.5}\selectfont
Observable chunks (unrouted / routed): 3,443 / 3,464; 869 / 872; 3,398 / 3,410; 1,517 / 1,515. Verification, applied to the unseen corpora, uses $\theta=.30$ and embeds 20.0 / 36.1\% of chunks. With a fixed router, three diffusion seeds give CANDOR 30\,s WT $3.99\pm0.48$\% (mean $\pm$ s.d.); entries use seed 0.
\end{minipage}
\endgroup
\end{table*}

\textbf{Enrollment and inference.} Anchors average unit ECAPA embeddings and are then renormalized. Training uses two 10\,s windows per speaker, preferentially outside the current excerpt; evaluation uses up to twenty activity-ranked windows from the same session's reference channel. All enrolled methods receive identical anchors, so comparisons isolate where identity is read. Main inference uses 30\,s chunks, 3\,s overlap, and ten DPM-Solver++ steps~\cite{lu2022dpmsolver}; the context test uses 120\,s/10\,s on the same CANDOR sessions.

\textbf{Comparison and metrics.} Baselines are no alignment, the released refined-latent overlap-cosine rule, and enrolled waveform re-identification, which represents post-reconstruction assignment under the same enrollment; post-hoc policies replay the same unrouted output, while routing generates a separate output with the same initial seed. WT is the duration-weighted fraction of observable chunks (overlap included) assigned against a reference-envelope oracle, which compares both pairings of mean-centered 20\,ms log-RMS envelopes (floor $10^{-3}$) after one per-session lag search in $\pm200$\,ms on the first six chunks; chunks with a pairing gap below 0.10 are excluded from numerator and denominator, separately per arm. Baseline orientation is calibrated on the first observable chunk; routing is scored in the requested order.

Activity strata use the smaller-to-larger reference RMS ratio, with cutoffs 0.1, 0.3, and 0.6, requiring the larger RMS to reach 0.02; these are acoustic strata, not annotated dialogue acts. Quality uses speaker-embedding cosine and activity F1, under both the better per-chunk permutation and the fixed reference order.

\section{Results and ablations}
\label{sec:results}
\subsection{Long conversations retain their participants}
Table~\ref{tab:main}a establishes the main result: routing reduces CANDOR WT from 43.86\% to 4.08\%, a 90.7\% relative reduction. Under the \emph{same enrollment information}, it reduces error by 34.2\% relative to waveform re-identification. Its advantage also reaches the most affected sessions: worst-session WT is 17.14\%, versus 42.72\% for waveform re-identification.

Longer context complements identity control rather than replacing it. Quadrupling the chunk length improves the unaligned baseline, while routing reaches 1.89\% WT against 4.72\% for enrolled re-identification, a 60.0\% relative reduction, so both context lengths favor the same identity interface.

\begin{figure}[t]
\centering
\includegraphics[width=\columnwidth]{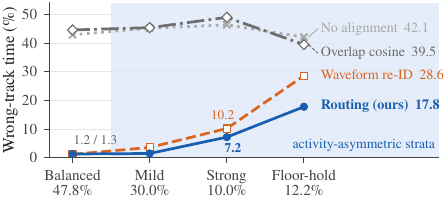}
\caption{\textbf{The gain is concentrated where one participant holds the floor.} Four recovery pipelines on CANDOR; tick labels give each stratum's share of speech-bearing chunk duration, and the shaded strata are activity-asymmetric.}
\label{fig:activity}
\end{figure}

Figure~\ref{fig:activity} locates the gain along the conversation's speaking--listening balance and states when routing is the method of choice. In balanced chunks, routing and waveform re-identification coincide (1.3\% versus 1.2\% WT); the difference arises in activity-asymmetric speech, where re-identification rises to 10.2\% and 28.6\% while routing holds it to 7.2\% and 17.8\%. Overlap matching remains near the unaligned level across strata. The mechanism is \emph{where} identity is read. Routing matches on the coarse latent streams with learned speech gating, so a near-silent listener stream contributes little to the decision; waveform re-identification must embed the decoded listener track, whose low-energy content yields unstable scores. In floor-holding chunks, 20.8\% of waveform decisions and 28.1\% of overlap-cosine decisions have margins below 0.05, versus 0.8\% of each in balanced chunks. Swap inheritance lets low-margin routing decisions keep the preceding order, so a listener's silence does not reset the convention. Routing is therefore the method of choice for long, activity-asymmetric conversations, which account for 52.2\% of speech-bearing duration on CANDOR; when both participants stay active, decoded-waveform evidence is already strong and the two approaches agree.

\subsection{Preserve the speech, restore its ownership}
Table~\ref{tab:main}b distinguishes reconstruction from participant assignment within the same evaluation columns. Under the better per-chunk permutation, speaker similarity and activity F1 differ by at most 0.004 across the four settings. Under the fixed convention, CANDOR activity F1 increases from 0.464 to 0.839, and from 0.471 to 0.851 with longer context. The CANDOR output/reference RMS ratio is stable at 0.497 versus 0.501. Routing changes who owns the speech, not what the speech sounds like.

\begin{table*}[t]
\centering
\caption{\textbf{Tests of the identity interface.} (a) CANDOR, 20 sessions / 8.33\,h, anchors fixed; speaker similarity and activity F1 are scored in the fixed reference order. (b) Same architecture and recipe; both use the same 480 development clips and 41+41 Seamless dyads. (c) Controlled interventions on routing decisions and on decoded audio; audio exchange is measured with swap inheritance disabled.}
\label{tab:tests}
\begingroup
\fontsize{8.2}{9.6}\selectfont
\setlength{\tabcolsep}{2pt}
% Left column (a)+(b) and right column (c) have matching natural heights; the panel grids share one row pitch.
\begin{minipage}[t]{.520\textwidth}
\settowidth{\lblwa}{Coarse values + diffusion}
\setlength{\colwa}{\dimexpr(\linewidth-\lblwa-8\tabcolsep)/4\relax}
\panel{(a) Where and how identity enters}\par\vspace{2pt}
\begin{tabular}{@{}>{\raggedright\arraybackslash}p{\lblwa} *{3}{>{\centering\arraybackslash}p{\colwa}}@{}}
\toprule
Intervention & \meta{Weights} & Spk.\,$\uparrow$ & F1\,$\uparrow$\\
\midrule
Coarse values + diffusion & \meta{adapted} & .429 & .689\\
Coarse-value adaptation   & \meta{adapted} & .489 & .782\\
\rowcolor{oursbg}
\textbf{\color{ours}Permutation only} & \meta{frozen} & \textbf{.531} & \textbf{.804}\\
\bottomrule
\end{tabular}
\par\vspace{6pt}
\panel{(b) Replace the training corpus}\par\vspace{2pt}
\begin{tabular}{@{}>{\raggedright\arraybackslash}p{\lblwa} *{4}{>{\centering\arraybackslash}p{\colwa}}@{}}
\toprule
Training source & \meta{Examples} & Dev.\,$\uparrow$ & Similar\,$\downarrow$ & Distinct\,$\downarrow$\\
\midrule
\rowcolor{oursbg}
\textbf{\color{ours}CANDOR} & \meta{29,958} & 94.17 & 9.19 & 2.46\\
Seamless improvised          & \meta{27,211} & 92.50 & 8.24 & 3.16\\
\bottomrule
\end{tabular}
\end{minipage}\hfill
\begin{minipage}[t]{.449\textwidth}
\settowidth{\lblwc}{Reverse anchors: non-tied decisions}
\setlength{\colwc}{\dimexpr(\linewidth-\lblwc-4\tabcolsep)/2\relax}
\panel{(c) Intervene on the requested identity}\par\vspace{2pt}
\begin{tabular}{@{}l *{2}{>{\raggedleft\arraybackslash}p{\colwc}}@{}}
\toprule
Intervention / response & Count & Rate (\%)\\
\midrule
\grp{Routing decision}\\
Oracle coarse order: track 0 = A    & 197/200 & \rate{98.5}\\
Half swap: first half unchanged     & 94/94   & \rate{100.0}\\
Half swap: second half exchanged    & 91/94   & \rate{96.8}\\
Reverse anchors: non-tied decisions & 9,516/9,516 & \rate{100.0}\\
\addlinespace[1pt]
\grp{Decoded audio exchange, matched-noise probe}\\
Distinct voices (20 dyads)          & 127/138 & \rate{92.0}\\
Highest activity-quality quartile   & 68/69 & \rate{98.6}\\
\bottomrule
\end{tabular}
\end{minipage}
\endgroup
\end{table*}

\begin{figure*}[t]
\centering
\begin{minipage}[t]{.486\textwidth}
\centering
\includegraphics[width=\linewidth]{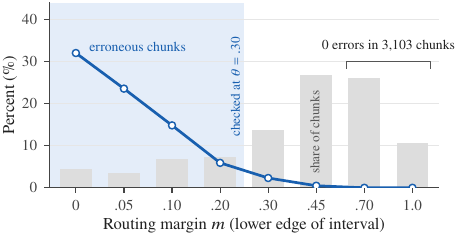}\\[-1pt]
\textbf{(a) Assignment risk and sample support}
\end{minipage}\hfill
\begin{minipage}[t]{.486\textwidth}
\centering
\includegraphics[width=\linewidth]{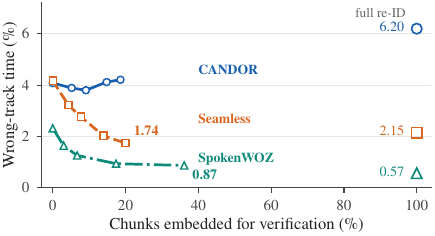}\\[-1pt]
\textbf{(b) Accuracy against recognition cost}
\end{minipage}
\caption{\textbf{Use extra recognition where it is informative.} (a) Eight margin intervals over 8,389 observable chunks; the line is erroneous-chunk percentage, grey bars give sample share, and the shaded intervals are those checked at the default $\theta=.30$. The first interval includes swap inheritance. (b) Routing plus verification at five shared thresholds ($\theta=0$ to $.30$, left to right), plotted against the share of chunks embedded; labels give WT at $\theta=.30$. Open markers are full waveform re-ID at 100\% coverage.}
\label{fig:verification}
\end{figure*}

\subsection{Why route before diffusion?}
\label{sec:ablation}
\textbf{Injection site and intervention type.} Table~\ref{tab:tests}a asks where identity should enter, with the CANDOR subset and the anchors fixed. Permutation routing is the only one of the three anchored alternatives that leaves every released weight frozen, and it also reaches the highest fixed-track speaker similarity (.531) and activity F1 (.804); anchor-conditioned diffusion adaptation reaches only .429 and .689. Upstream assignment is the effective control point, and its lightest form keeps the generator intact.

\textbf{Controlled identity changes.} Table~\ref{tab:tests}c tests whether coarse ordering actually controls the frozen generator. Oracle-ordering the coarse streams places speaker A on track 0 in 197 of 200 development clips. In the half-swap intervention, the unaltered first half retains its order in all 94 clips, while the second half follows the imposed exchange in 91. A local intervention produces a local identity response: this is the operational basis for assigning participants before diffusion.

Reversing the enrollment order tests the control the interface offers at two levels. At the decision level, all 9,516 distinct routing decisions in the reported configurations have nonzero margins and reverse exactly. At the audio level, the decoded tracks follow the exchange in 92.0\% of distinct-voice chunks and 98.6\% of chunks in the highest activity-quality quartile of the matched-noise probe.

\textbf{Training source.} Table~\ref{tab:tests}b replaces CANDOR with Seamless improvised training while retaining architecture, recipe, and evaluation clips. Development accuracy is 92.50\% versus 94.17\%, and matched similar-/distinct-voice errors stay within one percentage point: the latent-to-speaker correspondence is a property of the interface, not of one training corpus. All main results use the CANDOR checkpoint.

\subsection{Transfer at a fraction of the recognition cost}
\textbf{The same router, three recording domains.} The same CANDOR-trained router reduces WT to 4.17\% on Seamless and 2.32\% on SpokenWOZ, versus 18.66\% and 13.39\% without alignment (Table~\ref{tab:main}a), and raises fixed-track activity F1 from 0.546 to 0.853 and from 0.424 to 0.911 while best-permutation quality is retained (Table~\ref{tab:main}b).

\textbf{The margin identifies where to spend recognition.} Across all three corpora, the routing margin separates high-risk assignments from confidently routed speech (Fig.~\ref{fig:verification}a). The lowest-margin 22.3\% of the 8,389 observable chunks, exactly the region checked at the default $\theta=0.30$, contain 89.2\% of routing errors; 3,103 chunks with margins at least 0.70 have no observed errors.

\textbf{One operating point, a fraction of the recognition cost.} Figure~\ref{fig:verification}b places the same five thresholds on all three corpora against the share of output embedded; $\theta=0.30$ gives the last row of Table~\ref{tab:main}a. Seamless reaches 1.74\% WT while embedding 20.0\% of chunks, below the 2.15\% of full waveform re-identification at 100\% coverage; SpokenWOZ reaches 0.87\% while embedding 36.1\%, about a third of the recognition that full re-identification spends. On in-domain CANDOR, WT stays within 3.8--4.2\% at every threshold, below the 6.20\% of full re-identification.

Across the 360 Seamless dyads, the advantage holds in every speaker-similarity quartile: duration-pooled verified WT is 1.15/1.35/1.88/2.78\%, versus 2.04/1.41/2.44/2.80\% for full waveform re-identification.

\raggedbottom % the final text page ends with the conclusion; no stretch to distribute
\section{Conclusion}
Speaker-anchored routing makes persistent participants part of full-duplex dialogue recovery. On 25-minute conversations it cuts wrong-track time from 43.9\% to 4.1\% with 1.84\,M added parameters, 2.8\% added runtime, and a frozen generator. Its advantage is largest exactly where one participant holds the floor, and on face-to-face and telephone recordings its margin tells where extra recognition is worth spending: 20--36\% of the output instead of all of it. Matching coarse streams before diffusion changes who owns the speech, not what the speech sounds like. Full-duplex dialogue changes who is speaking, not who the participants are; preserving this distinction turns recovered voices into reliable conversation tracks.

\clearpage
\section{Compliance with Ethical Standards}
This research study was conducted retrospectively using human subject data made publicly available for research by their providers: CANDOR~\cite{reece2023candor}, Seamless Interaction~\cite{agrawal2025seamless}, SpokenWOZ~\cite{si2023spokenwoz}, and MUSAN~\cite{snyder2015musan}. Ethical approval was not required as confirmed by the licenses and terms of use attached to these data. No new recordings were collected, and the data were used only for research under those terms.

\nobalance
\bibliographystyle{IEEEbib}
\bibliography{references}
\end{document}